\documentclass{aa} 
\usepackage{txfonts}
\usepackage{natbib,graphicx,amssymb}

\begin{document}

\title{MgS in detached shells around carbon stars\thanks{based on
    observations obtained with ISO, an ESA project with instruments
    funded by ESA member states (especially the PI countries: France,
    Germany, the Netherlands and the United Kingdom) with the
    participation of ISAS and NASA}}

\subtitle{Mining the mass-loss history}

\author{
  S. Hony\inst{1} \and
  J. Bouwman\inst{2}
}

\institute{
  RSSD-ESA/ESTEC, PO Box 299, NL-2200 AG Noordwijk, The Netherlands
   \and
  CEA, DSM, DAPNIA, Service d'Astrophysique, CEA Saclay, F-91191
  Gif-sur-Yvette Cedex, France
}
 
\offprints{S. Hony, \email{shony@rssd.esa.int}}

\date{received \today; accepted date}

\abstract{We investigate the dust composition of detached shells
  around carbon stars, with a focus to understand the origin of the
  cool magnesium-sulfide (MgS) material around warm carbon stars,
  which has been detected around several of these objects
  \citep{2002A&A...390..533H}. We build a radiative transfer model of
  a carbon star surrounded by an expanding detached shell of dust. The
  shell contains amorphous carbon grains and MgS grains. We find that
  a small fraction of MgS dust (2\% of the dust mass) can give a
  significant contribution to the IRAS 25~$\mu$m flux. However, the
  presence of MgS in the detached shell cannot be inferred from the
  IRAS broadband photometry alone but requires infrared spectroscopy.
  
  We apply the model to the detached-shell sources R~Scl and U~Cam,
  both exhibiting a cool MgS feature in their ISO/SWS spectra. We use
  the shell parameters derived for the molecular shell, using the CO
  submillimetre maps \citep{1999A&A...351L...1L,2001A&A...368..969S}.
  The models, with MgS grains located in the detached shell, explain
  the MgS grain temperature, as derived from their ISO spectra, very
  well.  This demonstrates that the MgS grains are located at the
  distance of the detached shell, which is a direct indication that
  these shells originate from a time when the stellar photosphere was
  already carbon-rich.  In the case of R~Scl, the IRAS photometry is
  simultaneously explained by the single shell model. In the case of
  U~Cam, the IRAS photometry is under predicted, pointing to a
  contribution from cooler dust located even further away from the
  star than the molecular shell.
  
  We present a simple diagnostic to constrain the distance of the
  shell using the profile of the MgS emission feature. The emission
  feature shifts to longer wavelength with decreasing grain
  temperature. One can therefore infer a temperature and a
  corresponding distance to the star from the observed profile.  Such
  a diagnostic might prove useful for future studies of such systems
  with SIRTF or SOFIA.
}

\maketitle

\section{Introduction}
\label{sec:introduction}
\begin{figure}
  \centering \includegraphics[clip,width=8.8cm]{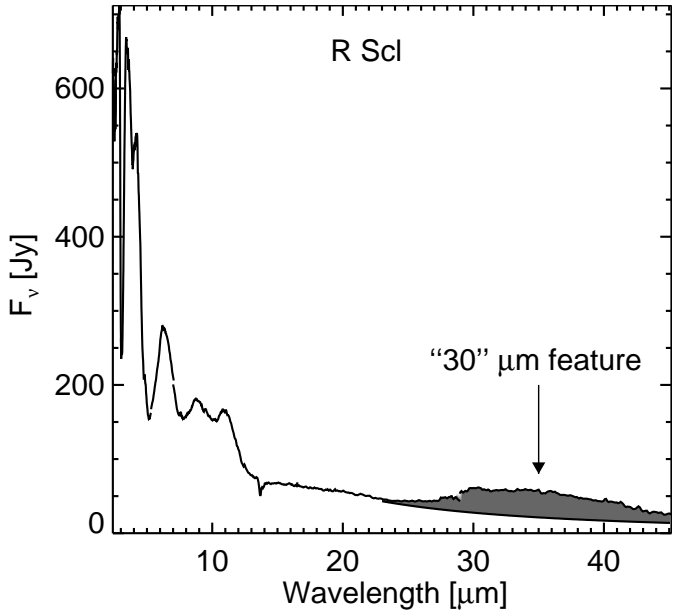}
  \caption{The ISO/SWS spectrum of R~Scl. An example of an optically
    bright carbon star that exhibits the ``30''~$\mu$m feature
    attributed to cool MgS grains. The ``30''~$\mu$m feature is shaded
    in grey. At shorter wavelengths there are several prominent
    photospheric, molecular absorption bands. The emission band at
    11.3~$\mu$m is attributed to silicon-carbide grains.}
  \label{fig:30um_feature}
\end{figure}
The so-called ``30''~$\mu$m emission feature is commonly detected in
the infrared (IR) spectra of carbon-rich (C-rich) evolved stars, from
low to intermediate mass-loss asymptotic giant branch stars (AGB)
through infrared carbon stars and post-AGB stars to planetary nebulae
\citep[e.g.][]{1981ApJ...248..195F,1993ais..conf...87O,1998Ap&SS.255..351Y,2002A&A...390..533H}.
The ``30''~$\mu$m feature, which exhibits itself as a prominent broad
emission excess in the IR spectrum, extending from $\sim$24 to
$\sim$45~$\mu$m, is commonly attributed to magnesium-sulfide (MgS)
dust grains \citep[e.g.][]{1985ApJ...290L..35G,1994ApJ...423L..71B}.
MgS grains are expected to form in C-rich environments
\citep{1978ApJ...219..230L,1999IAUS..191..279L}, but are chemically
unstable in oxidising, i.e. oxygen-rich, surroundings
\citep{1985ApJ...290L..41N}. The feature profile varies considerably
from source to source
\citep[e.g.][]{1985ApJ...290L..35G,2000tesa.conf....3W,2000ApJ...535..275H}.
\citet{2002A&A...390..533H} have presented a detailed spectral
comparison between the emission expected from MgS grains and the
profiles of the ``30''~$\mu$m emission feature found in the IR spectra
of C-rich evolved stars obtained with the Short Wavelength
Spectrograph \citep[SWS,][]{1996A&A...315L..49D} on board the Infrared
Space Observatory \citep[ISO,][]{1996A&A...315L..27K}.  Although there
are also some profile variations that can be attributed to grain-shape
variations, it was demonstrated that the main variations in the
appearance of the ``30''~$\mu$m feature between distinct sources can
be understood as a result of differences in the (average) MgS grain
temperature.  Conversely, one can derive the average grain temperature
from the observed profile.  One of the interesting findings of that
study was the discovery of several ``hot'', i.e. optically bright,
carbon stars that exhibit a cool MgS profile.  In
Fig.~\ref{fig:30um_feature} we show the spectrum of R~Scl. R~Scl is an
example of such an optically bright carbon star with a very prominent
``30''~$\mu$m feature.  One of the most likely explanations for this
phenomenon is the presence of MgS grains far from the star, whereas
there is a lack of MgS grains closer by.

Some optically bright carbon stars are known to possess a detached
shell of gas and dust around them
\citep[e.g.][]{1994A&A...281L...1W,2001A&A...368..969S}. In this
paper, we explore the effect that MgS grains located in such a
detached shell will have on the IR spectra and photometry of these
stars. We build a grid of radiative transfer models of a carbon star
surrounded by a detached dust shell which contains amorphous carbon
(a-C) and MgS grains. We compare the synthetic spectra from these
models with the IRAS photometry of known carbon stars. We further
apply the model to two specific carbon stars: R~Scl and U~Cam. The
detached shells around these sources are clearly resolved in molecular
line emission \citep{2001A&A...368..969S} and good quality ISO/SWS
spectra are available to compare with the synthetic spectra.  This
allows us to address the question on whether the MgS grains are
located at the same distance as these molecular shells.

\subsection{The origin of the dust far from the star }
\label{sec:origin-dust}
\begin{figure}
  \centering \includegraphics[clip,width=8.8cm]{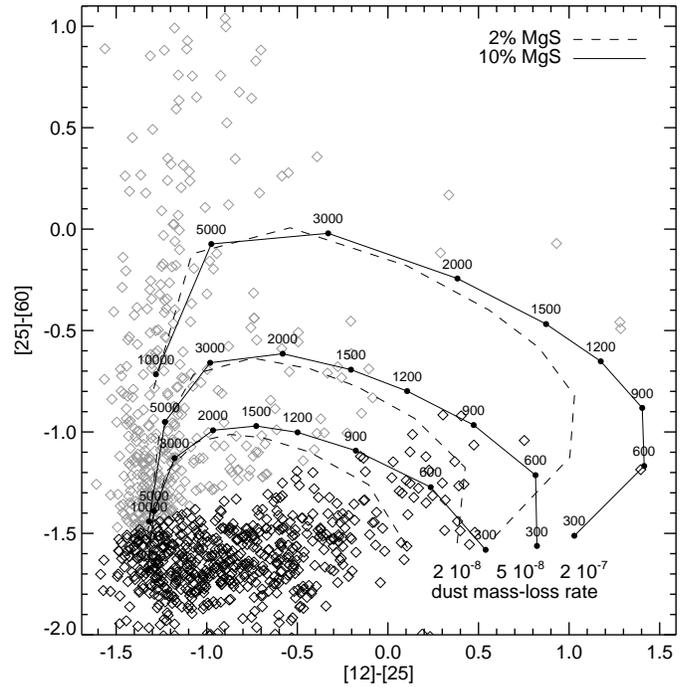}
  \caption{IRAS colour-colour diagram of galactic carbon
    stars (diamonds).  Stars that exhibit a single temperature SED are
    shown in black. Stars with a far-IR excess are grey.  We also show
    the tracks that the models follow in the diagram. The tracks are
    for a mass-loss burst lasting for 200 years.  The dust mass-loss
    rate increases from the bottom to the top curves
    (2\,10$^{-8}$-2\,10$^{-7}$ M$_{\odot}$/yr). The time elapsed since
    the burst is indicated in the figure. The difference between the
    dashed and the full lines is the relative amount of MgS. At the
    onset of the burst, when the dust is warmest, the contribution of
    MgS makes the tracks differ substantially. }
  \label{fig:iras2_color}
\end{figure}
Carbon stars are known to loose large amounts of envelope material
through a dust driven wind. The observational indications that this
mass loss may occur in a non-steady fashion are important for the
present study. Perhaps the most complete picture arises from the IRAS
colour-colour diagram \citep{1988A&A...194..125V} of C-rich evolved
stars (Fig.~\ref{fig:iras2_color}). This diagram shows the 60~$\mu$m
over 25~$\mu$m flux-ratio (expressed as a magnitudes difference)
versus the 25~$\mu$m over 12~$\mu$m flux-ratio based on the IRAS
photometry. A star's position in this diagram is thus a measure of the
shape of its IR spectral energy distribution (SED).  The diagram shows
that most of the carbon-star SEDs are dominated by a single
temperature. However, about one third of them show evidence of a
second, cooler dust component. This cooler dust component manifests
itself through a far-IR excess.

The common interpretation is that this excess is due to a detached
dust shell located away from the star
\citep[e.g.][]{1988A&A...196..173W,1993A&A...276..367Z}. This shell
originates from a previous epoch of mass loss, when the mass-loss rate
exceeded the present day mass-loss rate by several factors. Although
there has been some discussion in the literature on whether this
far-IR excess is actually related to the carbon stars
\citep{1995ApJ...445..415I,1996A&A...308..738E}, several of these
detached shells have now been directly detected. These shells have been
resolved either in the IR
\citep{1993ApJS...86..517Y,1994A&A...281L...1W,1995Ap&SS.224..495I,1996A&A...315L.221I,1997A&A...323..449I,2000AdSpR..25.2205I}
or in molecular lines in the submillimetre domain
\citep{1993PASJ...45..573Y,1996A&A...311..587O,1999A&A...351L...1L,2001A&A...368..969S}.
In a few cases, the optical light scattered by the material in these
shells has been directly detected
\citep{2000IAUS..177..425I,2001A&A...372..885G}.

The origin of these shells, i.e. the cause of the change in mass-loss
rate, is matter for debate.  The earliest suggestion was that
occurrence of these shell was closely related to the photospheric
composition of the AGB star. Specifically, it was suggested that these
shells arise as a result of the thermal pulse, that increased the
number ratio of C-atoms to O-atoms in the photosphere of the AGB star
from less than unity (O-rich) to larger than unity (C-rich)
\citep{1988A&A...196..173W,1988ApJ...334..362C}.  Later studies showed
that these shells may arise due to thermal pulses in general, as long
as the star is close to the tip of the AGB, i.e. with a luminosity
close to the critical luminosity
\citep[e.g.][]{1998A&A...335L...9S,2000A&A...357..180S}. Recent
numerical simulations of the momentum transfer in the dust driven
winds of AGB stars show that modulations in the frictional coupling
between the gas and the dust can also cause non-steady mass loss and
shell formation \citep{2001A&A...371..205S}.  These results have been
obtained for a C-rich gas and dust composition and it is at present
not clear whether the same mechanism applies to O-rich environments.

\subsection{The composition of the material in the shell }
\label{sec:intro-comp-shell}
Closely related to the question of the origin of these detached shells
is the question of their composition. The molecular and dust
composition in the shells is a tracer of the chemical and physical
conditions, in particular the C/O number ratio, in the atmosphere of
the star at the time the shell was formed. There are various indirect
evidences that these shells are C-rich.  \citet{1994A&A...288..551B}
find that the molecular line ratios for the various molecules are
indicative of a C-rich chemistry. From modelling of the SED, one also
tends to find that the opacities of C-rich dust agree better with the
observed far-IR excesses than O-rich dust
\citep[e.g.][]{1997A&A...321..605B}. However these models are not
conclusive and models with O-rich dust can also be fitted to the SEDs.
The possibility of detecting MgS grains in these shells is a
tantalising one, since this gives us a \emph{direct} handle on the
prevailing C/O ratio at the time the matter was ejected.  This is due
to the fact that the ``30''~$\mu$m feature, attributed to MgS grains,
is only found in C-rich environments and has, until now, not been
detected in O-rich environments.

The paper is organised as follows.  In
Sect.~\ref{sec:model-description} we present the description of the
model. The results of the model are presented and compared to IRAS
observations in Sect.~\ref{sec:results}. We apply the model to
\object{R~Sculptoris} and \object{U Camelopardalis} in
Sect.~\ref{sec:appl-sourc}. Both these sources have a ``30''~$\mu$m
feature in their ISO/SWS spectra and the detached gas shells have been
spatially resolved with molecular line observations.  This allows us
to test whether the molecules and dust are co-spatial.  Finally, in
Sect.~\ref{sec:disc--concl}, we summarise our findings and discuss
future prospects.

\section{Model description}
\label{sec:model-description}
Our model consist of a single spherical dust shell surrounding a
carbon star with a low current day mass-loss rate. The radiation of
the star is partially absorbed by the dust grains in the shell. The
absorbed light heats the dust grains, which will subsequently emit
light at IR wavelengths.

The optically visible C-star is represented by a Planck function of
2500~K. This temperature is representative for the ensemble of
optically bright carbon stars with cold MgS grains in the ISO/SWS
sample from \citet{2002A&A...390..533H}, where we use the
effective temperatures as given by \citet{2001A&A...369..178B} The
luminosity of the star is set to 10\,000~L$_{\odot}$, which
corresponds to a stellar radius of 535~R$_{\odot}$. Of course, the
whole system can simply be scaled to a different luminosity for the
central star by scaling all dimensions with the appropriate value.

\subsection{Shell parameters}
\label{sec:shell-parameters}
We assume a constant expansion velocity, with a typical value of
15~km/s. The expansion velocity is constant.  During the burst that
produced the shell, the mass-loss rate was constant.  Before and after
the burst the mass loss was negligible.  Due to the expansion, the
dimensions of the shell are related to the time since termination of
the burst as:
\begin{equation}
  \label{eq:1}
  R_\mathrm{in} = R_{\star}+v_\mathrm{exp}*t ~\mathrm{and}~
  R_\mathrm{out} = R_{in}+v_\mathrm{exp}*\Delta t, 
\end{equation}
where $R_\star$ is the radius of the star, $R_\mathrm{in}$ and
$R_\mathrm{out}$ are the inner and outer edges of the dust shell after
time $t$. $\Delta t$ is the duration of the high mass loss. We
calculate synthetic spectra for times after ejection ($t$) from 300 to
10\,000 year. For $t<300$ year the optical depth through the dust
shell is very large and these systems will look essentially like high
mass-loss carbon stars, instead of detached-shell sources.  For
$t>10\,000$ year the MgS grains are too cold to have any detectable
signature in the resulting SED. We vary $\Delta t$ between 200 to 1500
year. The corresponding values for the shell thickness (1\,10$^{16}$
$-$ 7\,10$^{16}$\,cm) bracket the thickness of the detached shells
derived in \citet{2001A&A...368..969S}

The gas density in the shell follows from $\rho_\mathrm{gas}(R) =
\dot{M}/(4\pi R^2 v_\mathrm{exp})$, where $\rho_\mathrm{gas}$ is the
gas density as a function of radius ($R$) and $\dot{M}$ is the
mass-loss rate. The dust density ($\rho_\mathrm{dust}$) is important
for the radiative transfer calculations. This value is the ratio of
the gas density to the gas-to-dust mass ratio ($\Psi$), which implies
that the models are determined by the dust mass-loss rate
($\dot{M}_\mathrm{dust}=\dot{M}/\Psi$). We calculate model spectra for
the dust mass-loss rate in the range 2\,10$^{-9}$ $-$ 2\,10$^{-7}$
M$_\odot$/yr. This translates to a total mass loss-rate in the range
5\,10$^{-7}$ $-$ 10$^{-4}$ M$_\odot$/yr, using a typical value of
$\Psi$=250. However, it is not clear, whether such a standard value is
always applicable (see also Sect.~\ref{sec:appl-sourc}).

\subsection{Dust parameters}
\label{sec:dust-parameters}
We use a-C and MgS as the composition of the dust. The optical
properties of a-C are taken from \citet{1993A&A...279..577P}. We use
an a-C grain-size distribution according to: $n(a) \propto a^{-3.5}
~\mathrm{with}~ 0.01 ~ \mu m \le a \le 1.0~\mu m$, where $n(a)$ is the
number density of grains with radius $a$. For the MgS grains we use
the same optical properties used to fit the observations of the C-rich
post-AGB star \object{HD 56126} \citep{2003A&A...402..211H}. These
properties are based on the calculated absorption cross-sections in
the IR, using the optical constants as published by
\citet{1994ApJ...423L..71B} We calculate the IR absorption
cross-sections for a continuous distribution of ellipsoids
shape-distribution. This distribution results in a resonance which
agrees best with the observed feature
\citep[e.g.][]{1994ApJ...423L..71B,1999A&A...345L..39S,2002A&A...390..533H}.
Unfortunately, the optical properties of MgS have not been measured
below 10~$\mu$m. We assume a constant absorption cross-section below
1~$\mu$m of $\pi\times$10$^{-4}$~$\mu$m$^2$ for a grain radius of
0.01~$\mu$m. We further assume a linear decrease in the absorption
cross-section to 0 from 1 to 2~$\mu$m and 0 between 2 and 10~$\mu$m.
The cross-section assumed here is identical to the one used to model
HD~56126.  The location of the dust around that star is well
determined from extensive IR imaging studies
\citep[e.g.][]{1998ApJ...492..603D,2000ApJ...544L.141J,2002ApJ...573..720K}
as the dust shell is clearly resolved. The assumed absorption
cross-section below 1~$\mu$m yields a MgS grain temperature, at the
location of the dust shell of HD~56126, which agrees well with its SWS
spectrum \citep{2003A&A...402..211H}.
  
We vary the relative proportion of MgS to a-C from 2 to 10 percent by
mass. A smaller fraction of MgS will not produce a significant
``30''~$\mu$m feature. A larger fraction of MgS is unlikely to be
present because the amount of MgS grains is limited by the abundance
of Mg and S atoms. For a solar composition gas, the maximum mass
contained in Mg+S pairs relative to the mass of the C-atoms is 20
percent. However, not all available Mg, S and C atoms are necessarily
be condensed into dust grains.

\subsection{Dust radiative transfer code}
\label{sec:dust-radi-transf}
We use the proprietary dust radiative transfer code \textsc{MODUST}.
This code solves the monochromatic radiative transfer equation, from
UV/optical to millimetre wavelength, in spherical geometry subject to
the constraint of radiative equilibrium, using a Feautrier type
solution method \citep{1964Feautrier,1978stat.book.....M}.  This
yields the temperature of the dust grains. The code allows to have
several different dust components of various grain sizes and shapes.
We refer to \citet{2000A&A...360..213B} and \citet{Bouwman_PhD} for a
description of techniques used in \textsc{MODUST}. \textsc{MODUST}
yields both the model SED and intensity maps. From the obtained model
spectra and intensity maps, it is straightforward to simulate
observations. For example, we simulate the synthetic SWS spectra by
overlaying the apertures as a function of SWS subband, as listed in
\citet{1996A&A...315L..49D}, on the intensity maps. IRAS broadband
fluxes are simulated by convolving the spectra with the IRAS
bandpasses \citep{1988IRAS..C......0B}.  We solve the transfer
equations at 36 wavelengths, on a logarithmically spaced wavelength
grid, from 0.2 to 1000~$\mu$m. This is sufficient for correctly
integrating over the wavelength, when performing the radiative
equilibrium calculation. These wavelengths also cover the entire SED.
In addition, we calculate the synthetic mid-IR spectrum from 2 to
45~$\mu$m, with linear wavelength steps of 0.5~$\mu$m, to compare with
the observed SWS spectra.

\section{Results}
\label{sec:results}
\begin{figure*}
  \centering
  \includegraphics[clip,width=18cm]{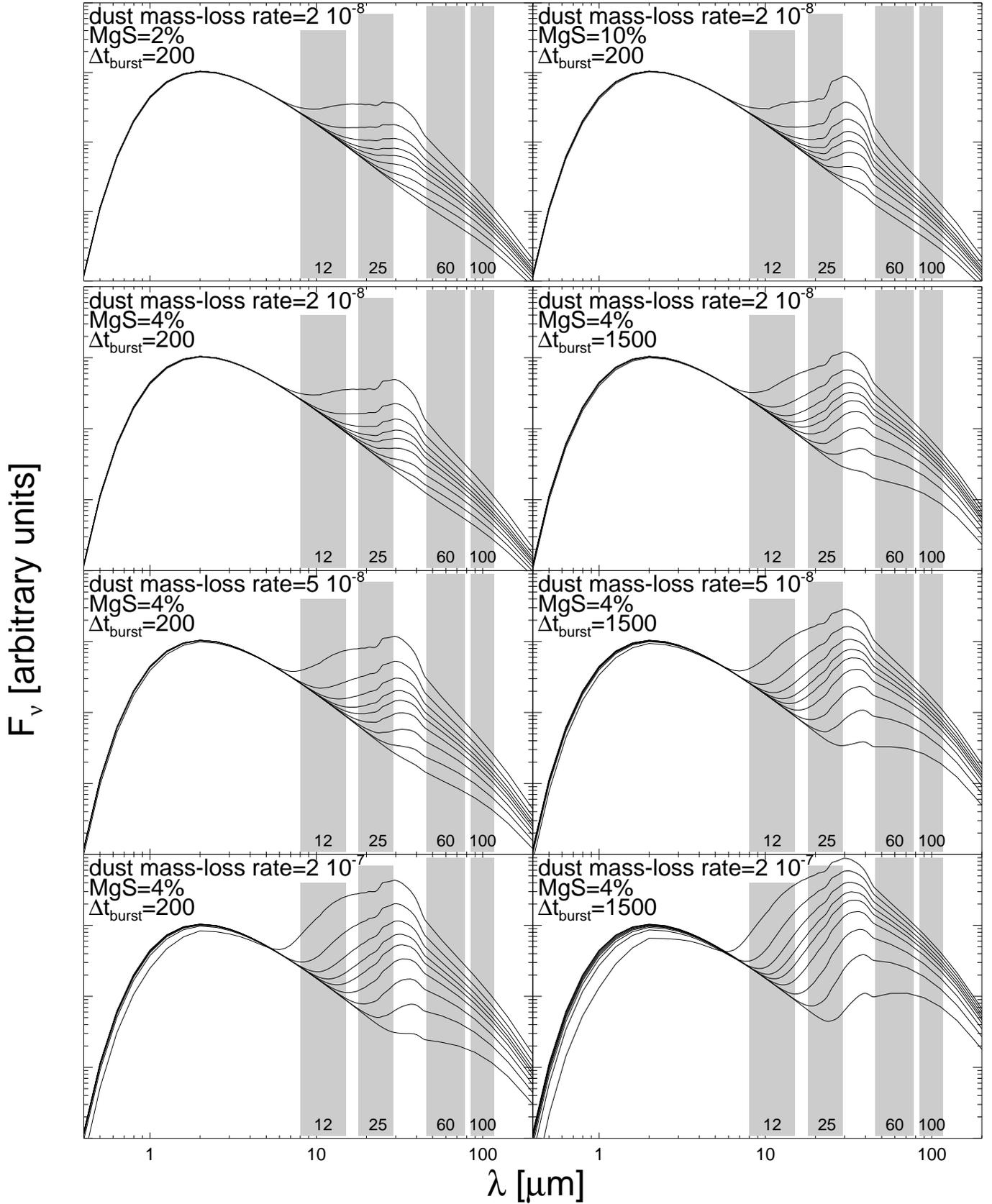}
  \caption{Synthetic SEDs from the expanding dust shell containing
    amorphous carbon and MgS grains.  In each panel, the curves from
    top to bottom represent the system 300, 600, 900, 1200, 1500,
    2000, 3000, 5000 and 10\,000 year after ejection of the shell. The
    top panels show the effect of varying the percentage of MgS
    grains.  In the bottom six panels, the mass-loss rate increases
    from top to bottom and the duration of the burst increases from
    left to right.  Also indicated are the wavelength regions that
    contribute to the different IRAS broadband filters.  The MgS
    grains are the cause of the ``30''~$\mu$m feature, the broad extra
    emission band between $\sim$23 and $\sim$45~$\mu$m. }
  \label{fig:seds}
\end{figure*}
In Fig.~\ref{fig:seds}, we show a subset of the SEDs that result from
our model. We show the main effects of varying the mass-loss rate, the
fraction of MgS grains and the duration of the burst.  We also
indicate which part of the SED contributes to the various IRAS
photometric filters.  There are a few general points that emerge from
the analysis.
\begin{enumerate}
\item The ``30''~$\mu$m feature is detectable even when present at a
  concentration of only 2 percent. This is especially true shortly
  after the ejection, when the MgS grains are relatively warm and the
  spectra clearly exhibit the characteristic onset at $\sim$26~$\mu$m.
\item MgS contributes considerably to the IRAS 25~$\mu$m flux during
  the first 2000 year when it is warm enough to emit within the IRAS
  25~$\mu$m bandpass.
\item The shell becomes quickly optically thin for the radiation of
  the central star and therefore the emission of MgS scales with the
  amount of MgS in the shell.
\item The 12~$\mu$m excess disappears within $\sim$750 year.
\item As expected, the IR excess increases with larger shell mass and
  is more pronounced for a short fierce burst than for a prolonged
  ejection at a lower mass-loss rate, even for the same amount of
  total material shed.
\end{enumerate}

\subsection{The IRAS colour-colour diagram}
\label{sec:iras-colour-colour}
\begin{figure}
  \centering
  \includegraphics[width=8.8cm,clip]{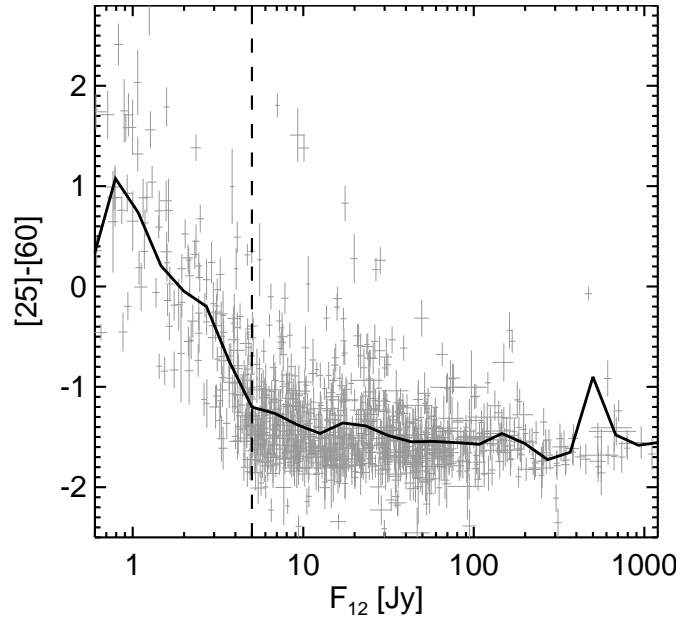}
  \caption{IRAS colour-magnitude diagram of galactic carbon stars. The
    thick black line shows the mean of the [25]-[60] colour as a
    function of 12~$\mu$m flux for the carbon stars (shown in grey).
    The sources with the largest 60~$\mu$m excess are all very faint
    at 12~$\mu$m (to the left of the dashed line). The photometry of
    these sources is probably affected by cirrus contamination or
    confusion due to the poor angular resolution of IRAS. }
  \label{fig:color_mag}
\end{figure}
In Fig.~\ref{fig:iras2_color} we compare the tracks as predicted from
our model with the IRAS colours of known galactic carbon stars.  The
diagram is constructed in the following way. We cross-correlate the
positions of the sources in the General Catalog of Galactic Carbon
Stars \citep{2001BaltA..10....1A} with the sources within
30$^{\prime\prime}$ in the IRAS point source catalogue
\citep{1988IRASP.C......0J}. This yields 3417 associations. From these
sources we only use those with a relative flux uncertainty less than
50 percent in the IRAS 12, 25 and 60~$\mu$m filter. In
Fig.~\ref{fig:iras2_color}, the maximum uncertainties are 0.55 and 0.6
and the average uncertainties are 0.11 and 0.15 in [12]-[25] and
[25]-[60], respectively.

The tracks explain the general features of the observed colour-colour
diagram well. The colours of the systems immediately after ejection of
the shell are similar to the colours of the infrared carbon stars,
because of the large optical depth in the shell. The expansion of the
shell causes the optical depth to rapidly decrease and the 12~$\mu$m
excess to diminish as the dust becomes cooler.  Therefore, the source
moves rapidly to the left in the diagram.  The stars spend by far the
longest time on the vertical track, at [12]-[25] $\approx$ -1.2,
while the 60~$\mu$m excess due to the cool dust slowly diminishes.
This explains the clustering of the stars in this region of the
diagram.  The effect of the ``30''~$\mu$m feature due to MgS grains
can only be noticed within 2000 years after ejection.  However, the
influence of MgS on the colours is not unique, because the effect of
the MgS grains in the colour-colour diagram can be mimicked by a
higher mass-loss rate or a longer burst duration.  This implies that,
due to the presence of MgS grains in these shells, and thus of the
``30'' $\mu$m feature, the mass in the shell may be overestimated,
when determined from the IRAS photometry alone.
\begin{figure}
  \centering
  \includegraphics[width=8.8cm,clip]{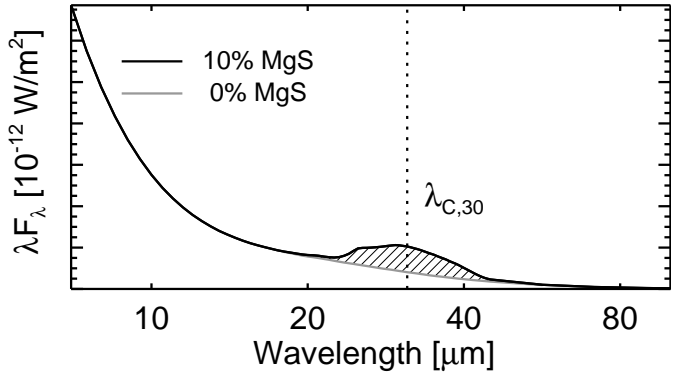}
  \caption{Illustration of the definition of the centroid position of
    the ``30''~$\mu$m feature. We shows two model spectra of
    a detached shell containing 0 and 10 percent MgS. The difference is
    the ``30''~$\mu$m feature due to the presence of the MgS grains.
    On this scale equal surface corresponds to equal energy. The
    centroid position is indicated by the dashed line, which divides
    the ``30''~$\mu$m feature into equal surface. }
  \label{fig:measure30}
\end{figure}

We note that there are several carbon stars in the IRAS database with
60~$\mu$m excesses larger than the ones shown in
Fig.~\ref{fig:iras2_color}.  We have not attempted to reproduce these
IRAS colours with our model.  It is likely that there is a sizeable
contamination of the far-IR fluxes among the carbon stars, as has been
argued by \citet{1995ApJ...445..415I}.  Specifically, we point out
that the largest IRAS excesses are found for the sources with low
12~$\mu$m fluxes, i.e. sources far away with a much larger chance for
confusion or cirrus contamination.  The carbon stars with IRAS colour
[25]-[60] $>$ 0 are almost exclusively found among sources with
F$_{12}$ $<$ 5 Jy (see Fig.~\ref{fig:color_mag}).  In this paper, we
are only concerned with the sources that have [25]-[60] $\lesssim$ 0.5
and clearly have a detached shell associated with them.  In order to
obtain 60 $\mu$m excesses in the range [25]-[60] $\simeq$ 1 or even
higher (see Fig.~\ref{fig:color_mag}), we have to increase the total
dust mass to 10$^{-4}$\,M$_\odot$, which in turn yields an estimated
mass-loss rate an order of magnitude higher than those derived for the
best studied cases.  It is not clear whether the same detached-shell
model can be invoked to explain such extreme excesses.

\subsection{MgS as a temperature/distance diagnostic}
\label{sec:mgs-as-temperature}
\begin{figure}
  \centering
  \includegraphics[width=8.8cm,clip]{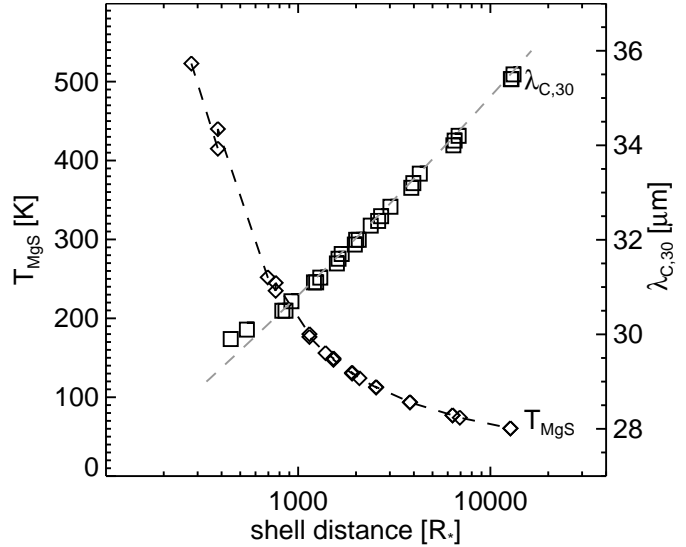}
  \caption{MgS temperature and the centroid position of the ``30''~$\mu$m
    feature as a function of distance to the star. The diamonds show
    the MgS temperature at the inner edges of the detached dust shell.
    The squares represent the centroid position of the ``30''~$\mu$m
    emission feature versus the typical distance of the shell, which
    is defined as the inner radius plus one quarter of the shell
    thickness. The grey dashed line shows a simple analytic
    representation of the relation between the distance and the
    centroid position (see Eq.~(\ref{eq:centr_dist})).}
  \label{fig:diagnostic}
\end{figure}
Because the shell becomes rapidly optically thin, we can give an
approximate temperature of the MgS grains as a function of shell
radius and effective temperature of the star. The effects of the
expansion of the shell are shown in Fig.~\ref{fig:diagnostic}. We show
two curves. The diamonds show the dependence of temperature on
distance. The squares show how this temperature translates into a
shifting of the ``30''~$\mu$m feature (see also Fig.~\ref{fig:seds}).
The shift is expressed in the centroid position of the MgS feature,
i.e. the wavelength which divides the feature in equal energy halves,
see also Fig.~\ref{fig:measure30}. This quantity is in general more
easily derived from the spectra and is also a better diagnostic of the
temperature than the peak position. For the distance of the shell we
take a ``typical'' distance, defined as the inner radius plus one
quarter of the shell thickness. This is to account for the fact that
the dust on the inside of the shell is hotter and thus brighter than
on the outside. The points shown in Fig.~\ref{fig:diagnostic} are for
shell thickness between 250 and 2000 $R_{\star}$, which corresponds to
a burst duration from 200 to 1500 years. An approximate power-law
expression (valid for 30 $<$ $\lambda_\mathrm{C,30}$ $<$ 36~$\mu$m) is
given by:
\begin{equation}
  \label{eq:centr_dist}
  \frac{R_\mathrm{shell}}{R_\star} =
  \left({\frac{\lambda_\mathrm{C,30}}{21~\mu m}}\right)^{18} \times
  \left(\frac{T_\star}{2500~K}\right)^{2},
\end{equation}
where $R_\mathrm{shell}$ is the typical distance of the shell.  Note
that, in physical terms, the shape of the emission feature is a
weighted sum of Planck functions, depending on the MgS grain
temperatures, convolved with the wavelength dependent MgS absorption
cross-section.  Therefore, the simple power-law expression and the
constants in the first term of Eq.~(\ref{eq:centr_dist}) lack a direct
physical interpretation.  With optical photometry of the central star,
both $T_{\star}$ and $R_\mathrm{shell}$ (in arcsec) can be determined
independent of distance, as the angular size of the star ($R_\star$ in
arcsec) is set by $T_\star$ and the photometry alone.  How practical
Eq.~(\ref{eq:centr_dist}) will be in reality remains to be
demonstrated, because it is known that there are also some differences
in the shape of ``30''~$\mu$m feature between sources that are not
caused by temperature but are more likely due to the grain-shape
distribution \citep{2002A&A...390..533H}. Such a difference can shift
$\lambda_\mathrm{C,30}$ by up to 2.5~$\mu$m: \emph{a factor 4 in shell
  radius!} Because the main deviation in the profile is a 26~$\mu$m
excess, the tendency will be to find a lower value of
$\lambda_\mathrm{C,30}$ and therefore derive a too small shell radius.

\section{Application to sources with ISO/SWS spectra }
\label{sec:appl-sourc}
In the previous section we have shown that the presence of MgS in the
detached shells can have a significant effect on the IRAS~25~$\mu$m
photometry. However, the presence of MgS grains in the shell cannot be
inferred from the IRAS photometry alone. In this section we focus on
those warm carbon stars that show a ``30''~$\mu$m feature in their SWS
spectrum.

\subsection{Application to R~Sculptoris}
\label{sec:application-r-scl}
\begin{figure*}
  \includegraphics[clip,width=12cm]{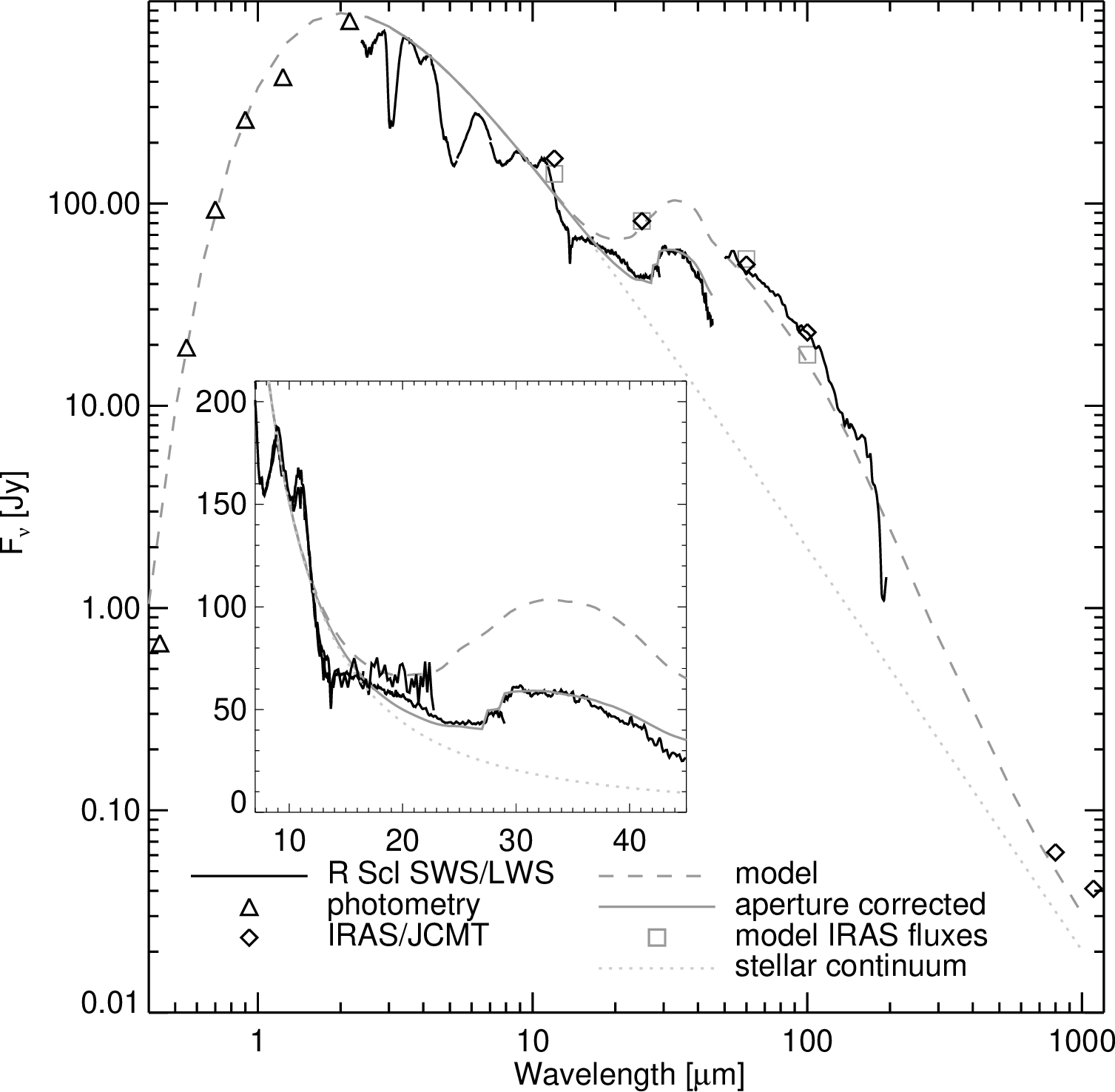}
  \caption{The detached-shell model compared to the observations of
    R~Scl in black. The optical photometry is shown in triangles
    \citep{1966CoLPL...4...99J}. Between 2 and 45~$\mu$m we show the
    SWS spectrum \citep{2002A&A...390..533H}; between 45 and
    200~$\mu$m we show the LWS spectrum. The diamonds represent the
    IRAS measurements and the submillimetre points from
    \citet{1998MNRAS.296..545B}. The dashed gray line is the model
    SED, taking identical parameters for the dust shell as derived for
    the CO shell by \citet{2001A&A...368..969S}. The full gray line
    shows the model spectrum after correction for the limited size of
    the apertures of the SWS spectrograph.  The dotted line represents
    the photospheric continuum.  We also show the synthetic IRAS
    fluxes as derived from the model SED. The 25 and 60~$\mu$m IRAS
    fluxes are in excellent agreement.  The predicted flux level at
    longer wavelengths is slightly too low.  The model explains the
    observations very well. In the inset, we show the fit to the
    ``30''~$\mu$m feature on a linear scale. We also show the IRAS/LRS
    spectrum, which starts to deviate from the SWS spectrum at the
    wavelength, where the model predicts a difference, due to the
    different beam-size of IRAS and SWS.}
  \label{fig:fit_r_scl}
\end{figure*}
\begin{figure}
  \centering \includegraphics[clip,width=7cm]{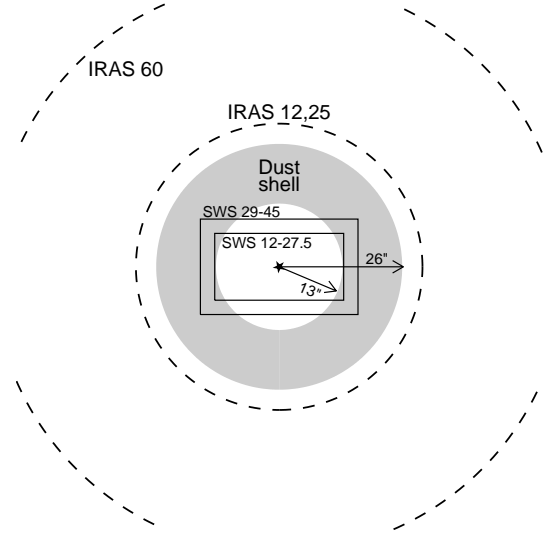}
  \caption{The angular dimensions of the circumstellar shell of R~Scl
    compared with the regions probed by ISO/SWS and IRAS. The black
    rectangles indicate the dimensions of the two most relevant SWS
    apertures. The dashed circles give an indication of the regions
    corresponding to IRAS 12, 25 and 60~$\mu$m point source
    measurements. The circle for the IRAS 100~$\mu$m filter lies
    outside the figure. The shell extends from 13$^{\prime\prime}$ to
    26$^{\prime\prime}$, which corresponds to 7\,10$^{16}$ $-$
    14\,10$^{16}$\,cm at a distance of 360~pc.  The complete shell is
    probed by the IRAS measurements. The SWS instrument detects only a
    part of the shell radiation. Of course, the back and front side of
    the spherical shell do contribute to the SWS flux.}
  \label{fig:r_scl_config}
\end{figure}
\begin{table}
  \begin{tabular}{@{\ }l l l@{\ }l l@{\ } l l l@{\ } l l@{\ }}
    \hline
    \hline
    & 
    &
    \multicolumn{5}{c}{Gas shell$^a$}&
    &
    \multicolumn{2}{c}{Dust shell}
    \\
    \cline{3-7}
    \cline{9-10}
    \\
    Source& 
    d& 
    $R_\mathrm{in}$&
    $R_\mathrm{out}$& 
    $R_\mathrm{in}$&
    $R_\mathrm{out}$& 
    $\dot{M}$&
    &
    $\dot{M}_\mathrm{dust}$& 
    MgS
    \\
    &
    [pc]&
    \multicolumn{2}{c}{[$^{\prime\prime}$]}& 
    \multicolumn{2}{c}{[10$^{16}$\,cm]}& 
    \multicolumn{3}{c}{[M$_\odot$/yr]}&
    \%
    \\
    \hline
    R Scl & 360 & 13 & 26 & 7 & 14  & 4\,10$^{-6}$ && 8\,10$^{-9}$ & 4 \\
    U Cam$^b$ & 500 & 7 & 9 & 5.5 & 6.6 & 9\,10$^{-6}$ && 1.6\,10$^{-8}$ & 4 \\
    \hline
  \end{tabular}
  \caption{Shell parameters of R~Scl and U~Cam. $^a$Taken from
    \citet{2001A&A...368..969S}. $^b$The dust parameters for U Cam
    only concern the dust located inside the SWS aperture (see text
    for details).} 
  \label{tab:model_details}
\end{table}
We first apply the model to R~Scl. This star is among the best studied
optically bright carbon stars with a far-IR excess. Its detached shell
has been resolved in molecular lines emission
\citep{1996A&A...311..587O} and scattered light from the detached
shells is detected \citep{2001A&A...372..885G}. Furthermore, R~Scl has
the strongest ``30''~$\mu$m feature of the optically bright carbon
stars observed with SWS.

For the parameters of the dust shell (see
Table~\ref{tab:model_details}), we take the inner and outer edge of
the shell of the detached gas shell as listed by \citet[][ their
Table~6]{2001A&A...368..969S}.  These parameters are derived from
mapping observations of the molecular CO emission, in which the shell
is clearly resolved. We use the same distance to the star as these
authors do, to facilitate a direct comparison of the parameters, which
they derive and ours. With the inner and outer edge of the shell
fixed, we have only two free parameters: the dust mass-loss rate and
the mass fraction of MgS. We vary these parameters and compare the
results with the IRAS and ISO observations.  The strength of the IR
continuum excess is set by the amount of amorphous carbon in the
shell, which scales linearly with the dust mass-loss rate. The
fraction of MgS influences the strength of the resultant ``30''~$\mu$m
feature.  The best agreement with the observations is found with 4
percent of MgS and a dust mass-loss rate of only
8\,10$^{-9}$\,M$_{\odot}$/year.  This best fit is shown in
Fig.~\ref{fig:fit_r_scl}.  We derive a total dust mass in the shell,
of 1.2\,10$^{-5}$\,M$_{\odot}$. The gas mass as determined by
\citet{2001A&A...368..969S} is 5.4\,10$^{-3}$\,M$_{\odot}$, which
yields for the gas-to-dust mass ratio: $\Psi$=450. The model
parameters are given in Table~\ref{tab:model_details}.

We show a sketch of the model, with the dimensions of the relevant SWS
apertures and the IRAS beam-sizes overlaid, in
Fig.~\ref{fig:r_scl_config}. The IRAS measurements sample the complete
detached shell, whereas the SWS spectrograph detects only a fraction
of the shell. The different regions probed by the various instruments
give rise to differences in the flux levels measured by these
instruments. The IRAS measurements sample the complete shell and
should therefore yield much higher flux level than SWS does.  This is
indeed as observed (see Fig.~\ref{fig:fit_r_scl}). A smaller
difference in the flux level, due to the size of the SWS apertures, is
also observed in the SWS data below 27.5~$\mu$m and above 29~$\mu$m.

The model is able to explain the strength of the MgS resonance within
the SWS aperture accurately. It simultaneously explains the much
stronger excess in the IRAS beam.  The presence of the MgS excess was
already suspected by \citet{1998MNRAS.296..545B} who found that the
IRAS 25 $\mu$m photometry was higher than expected on the basis of the
flux level of their Cooled Grating Spectrograph (CGS3) spectrum which
goes out to 23.5~$\mu$m. These authors speculate that the high IRAS
25~$\mu$m flux level might be due to a strong emission feature, at
wavelengths longer than $\simeq$ 24~$\mu$m, i.e.  the ``30''~$\mu$m
feature.  When we convolve the synthetic SED with the IRAS filter
profile, the IRAS 25~$\mu$m flux is indeed nicely reproduced by our
model.

The excellent agreement, between the simple detached-shell model and
the ISO \emph{and} IRAS data, demonstrates that the ``30''~$\mu$m
feature arises from the location of the detached shell. We conclude,
that MgS grains are present in the detached shell of R~Scl. We would
like to emphasise that we do not attempted to further optimise the
obtained model by varying the inner or outer edge of the dust shell.
We have to make a rather crude assumption for the absorption level of
the MgS grains below 1~$\mu$m (Sect.~\ref{sec:dust-parameters}).
Therefore, this simple model does not warrant such a detailed fitting
procedure.  However, the correspondence between the model and
observations, \emph{especially those probing different spatial
  scales}, demonstrates that the MgS emission arises from roughly the
same location as the molecular gas emission.
 
\subsection{Application to U Camelopardalis}
\label{sec:appl-u-camel}
\begin{figure*}
  \includegraphics[width=12cm,clip]{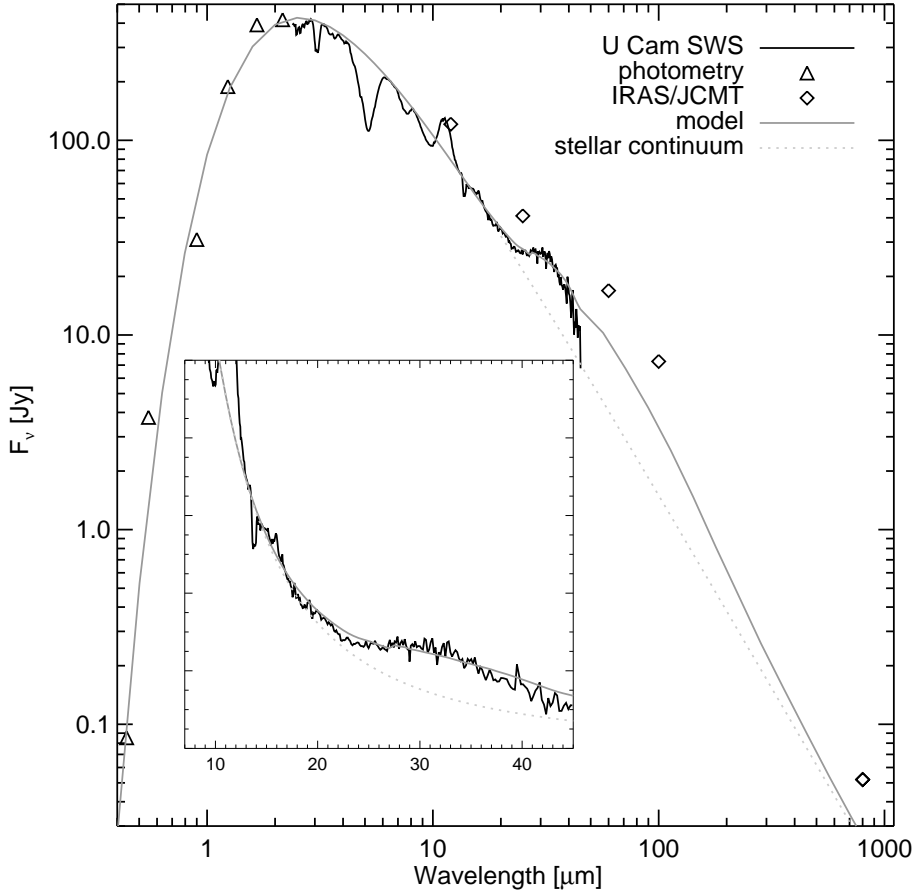}
  \caption{The single detached-shell model applied to U Cam. Symbols
    are the same as in Fig.~\ref{fig:fit_r_scl}. We use the inner and
    outer radius and the gas mass of the CO shell and a 4\% MgS
    fraction. The model that fits the SWS observations best does not
    reproduce the IRAS flux levels. Clearly, the single dust shell
    model is not applicable and the IRAS excess is predominantly due
    to more extended emission than the CO shell.}
  \label{fig:u_cam}
\end{figure*}
We also apply the model to U~Cam. Compared to R~Scl, this object has a
much thinner CO shell (Table~\ref{tab:model_details}).
\citet{1999A&A...351L...1L} measured an inner radius of
5.5\,10$^{16}$\,cm with a thickness of only 1.1\,10$^{16}$\,cm at a
distance of 500~pc. This corresponds to an outer radius of
9$^{\prime\prime}$ and the shell fits entirely within the largest SWS
aperture.  With those values for the inner and outer radius of the
shell the model cannot explain the IRAS and the ISO measurements
simultaneously (Fig.~\ref{fig:u_cam}). We find that the low IR flux
levels, as measured by ISO/SWS, place a stringent constraint on the
amount of dust located at the position of the CO shell. However, the
IRAS fluxes clearly indicate that there is more dust present in the
system.  This dust is probably located even further away from the
star. As can be seen in Fig.~\ref{fig:u_cam}, all the IRAS fluxes as
well as the 800~$\mu$m flux are under-predicted by the shell model.
When we increase the shell dust mass, in order to agree with the IRAS
25 and 60~$\mu$m flux levels, the longer wavelength fluxes are still
very much under predicted by the relatively warm dust in the model.
This indicates that the system contains cooler dust, which is probably
located further away.  The latter conclusion is in qualitative
agreement with the fact that the interferometric measurements of
\citet{1999A&A...351L...1L} retrieved about half of the total CO
emission, measured with single dish observations
\citep{1998A&AS..130....1N}. This is due to a more extended component
of CO emission.
  
The situation in the case of U~Cam is clearly more complex than for
R~Scl. The observations can not be explained by a single dust shell.
The far-IR fluxes levels are dominated by dust located outside the
detected CO shell. However, we stress that there is MgS present in the
CO shell.  The ISO/SWS instrument, which only probes the inner
regions, detects a ``30''~$\mu$m feature. The temperature of the MgS,
as derived from the centroid position of the ``30''~$\mu$m feature in
the ISO observation, yields a distance for the dust shell of
5.5\,10$^{16}$~cm (7.3$^{\prime\prime}$ at a distance of 500~pc).
This distance is identical to the size of the detached CO shell. Using
the gas mass of the shell of 1.4\,10$^{-3}$\,M$_{\odot}$ as given by
\citet{2001A&A...368..969S} and a dust mass of
2.4\,10$^{-6}$\,M$_{\odot}$ from the best fitting model for the
ISO/SWS data, we find a gas-to-dust mass ratio of $\Psi$=580.

We would like to point out that, in the two cases that we study here,
the derived gas-to-dust mass ratio in the shell is relatively high
(450 for R~Scl and 580 for U~Cam). These numbers are much higher than
the values between 200 and 250, which are commonly used for modelling
carbon-rich AGB and post-AGB stars
\citep[e.g.][]{1986ApJ...303..327J,1997ApJ...482..897M}. However, the
numbers we derive are in relatively good agreement with the high
values of \citet{2001A&A...368..969S}. One should clearly be
cautious in applying a ``standard'' value, when determining the total
shell masses for such systems based on the IR excess alone.

\subsection{Other warm carbon stars with the ``30''~$\mu$m
  feature }
\label{sec:other-warm-c-stars}
\begin{table}
  \centering
  \begin{tabular}{l l c c}
    \hline
    \hline
    \multicolumn{1}{c}{Source}& 
    \multicolumn{1}{c}{${\mathrm{m}_\mathrm{bol}}^{1}$}&
    \multicolumn{1}{c}{${\lambda_\mathrm{C,30}}^{2}$ [$\mu$m]}& 
    \multicolumn{1}{c}{R$_{shell}$ [$^{\prime\prime}$]}
    \\
    \hline
    \object{W Ori} & 2.87 & 31.3 & 8\\
    \object{S Cep} & 3.15$^\dagger$ & 31.2 & 6\\
    \object{T Dra} & 4.94 & 30.2 & 1.6\\
    \object{V Cyg} & 3.24$^\dagger$ & 30.5 & 4\\
    \object{RU Vir}& 4.24$^\dagger$ & 30.4 & 2\\
    \hline
    \hline
  \end{tabular}
  \caption{Estimated detached-shell radii (R$_\mathrm{shell}$) of warm
    carbon stars exhibiting a cool MgS feature in their SWS spectra,
    based on the relation in Eq.~(\ref{eq:centr_dist}) between the
    ``30'' $\mu$m feature centroid position and the size of the
    shell. $^1$Bolometric magnitude from 
    \citet{2002A&A...390..967B}. $^2$Centroid positions taken from
    \citet{2002A&A...390..533H}. $^\dagger$Average of the listed
    values.}
  \label{tab:positions}
\end{table}
There are a few more warm and optically bright carbon stars that
exhibit a cool MgS feature in their SWS spectra. We do not model these
stars here, mostly because of the lack of suitable molecular data to
compare with. However, we do give a crude estimate of their shell
radii. We calculate the angular radius of the star times the effective
temperature squared from the bolometric magnitude given by
\citet{2002A&A...390..967B}, using:
\begin{eqnarray}
  \label{eq:3}
  m_{\mathrm{bol},\star} &=& M_{\mathrm{bol},\star}
  -2.5\log_{10}\left(\left(\frac{10}{d}\right)^2\right) + 
  M_{\mathrm{bol},\odot} - M_{\mathrm{bol},\odot} \nonumber \\
  &=& M_{\mathrm{bol},\odot}
  -5\log_{10}\left(\left(\frac{R_\star}{R_\odot}\right)\left(\frac{T_\star}{T_\odot}\right)^2\left(\frac{10}{d}\right)\right)
  \Rightarrow \nonumber \\
  R_{\star,\mathrm{arcsec}}{T_\star}^2 &=&
  \left(\frac{\left(M_{\mathrm{bol},\odot} -
  m_{\mathrm{bol},\star}\right)}{5}\right)^{10} \frac{R_\odot}{10\,AU}
  {T_\odot}^2,
\end{eqnarray}
where $m_{\mathrm{bol}}$ and $M_{\mathrm{bol}}$ are the apparent and
absolute bolometric magnitude and $R_{\star,\mathrm{arcsec}}$ is the
angular radius of the star. The measured values of the apparent
bolometric magnitude and the centroid position of the MgS feature
together with Eq.~(\ref{eq:centr_dist} \& \ref{eq:3}) yields the
angular radius of the shell.  The sources and the estimated radii are
listed in Table~\ref{tab:positions}. These sources all have much
weaker ``30'' $\mu$m features than the feature observed in the SWS
spectrum of R~Scl, with a correspondingly lower signal-to-noise ratio.
At the same time, the estimated stellar continuum underlying the
``30''~$\mu$m feature becomes more influential on the derived feature
profile.  Based on this, we estimate an uncertainty in
$\lambda_\mathrm{C,30}$ of $\gtrsim$1~$\mu$m and in the derived shell
radii of at least a factor 2.

\section{Conclusions \& future prospects}
\label{sec:disc--concl}
With the models that we have presented, we now have a tool to
systematically investigate the influence of MgS on the IR spectra of
carbon stars with detached shells. We find that, in the case of R~Scl,
the dust (containing MgS) is co-spatial with the molecular gas. This
shows that the dust present in the shell is indicative of C-rich
chemistry and therefore \emph{the photosphere was C-rich at the time
  the shell was ejected}. This rules against a scenario in which the
shell is produced due to the transition from O-rich to C-rich. We find
that the shell around U~Cam contains MgS as well, again demonstrating
the C-based chemistry of the shell.

The prominence of the ``30''~$\mu$m feature and the broad wavelength
range over which the MgS grains emit make this material an excellent
tool for future studies of the circumstellar shells of carbon-rich
evolved stars. In particular, we point out that, because of the
temperature dependence of the MgS emission profile, some of the usual
ambiguity in interpreting SEDs can be lifted, as we can constrain
the location of the shell by means of the ``30''~$\mu$m feature. The
low-resolution long-wavelength mode of the IRS instrument
\citep{1998SPIE.3354.1192R} on board SIRTF \citep{2003SPIE.4850...17G}
will be sensitive enough to easily detect the MgS feature in carbon
stars located in the SMC or LMC. This will eliminate the distance
uncertainty, which usually troubles the derivation of the shell
parameters of galactic carbon stars and therefore allows a more
statistical approach for studying the occurrence rate and shell masses
of such detached shells.

Another tantalising prospective is offered by the FORECAST instrument
\citep{2000SPIE.4014...86K} for the SOFIA observatory
\citep{1997fisu.conf..201B}. With this camera it will be feasible to
directly image the shells at the wavelengths where the MgS is
emitting. The shells discussed here are all easily resolved with the
good angular resolution ($<$ 5$^{\prime\prime}$) of the large SOFIA
mirror. The several bandpasses in the ``30''~$\mu$m feature will allow
us to study the MgS abundance and temperature as a function of
distance to the carbon star to an unprecedented level of detail.
\begin{acknowledgements}
  The authors thank Frank Molster, Jacco van Loon and Rens Waters for
  interesting and very constructive discussions. We thank Coralie
  Neiner for careful reading of the manuscript.  JB acknowledges
  financial support from the EC-RTN on ``The Formation and Evolution
  of Young Stellar Clusters'' (RTN-1999-00436, HPRN-CT-2000-00155).
  This research has made use of the SIMBAD database, operated at CDS,
  Strasbourg, France.  This research has made use of NASA's
  Astrophysics Data System Bibliographic Services. This research has
  made use of the VizieR catalogue access tool, CDS, Strasbourg,
  France.
\end{acknowledgements}

\bibliographystyle{aa}
\bibliography{articles}
\end{document}